# Vicinity Effects of Field Free Point on the Relaxation Behavior of MNPs

Atakan Topcu[1], Asli Alpman[1,2], Mustafa Utkur[1,2], Emine Ulku Saritas[1,2,3]

[1]Department of Electrical and Electronics Engineering, Bilkent University, Ankara, Turkey
[2]National Magnetic Resonance Research Center (UMRAM), Bilkent University, Ankara, Turkey
[3]Neuroscience Program, Sabuncu Brain Research Center, Bilkent University, Ankara, Turkey
Email: atakan.topcu@ug.bilkent.edu.tr

***Abstract:*** In Magnetic Particle Imaging (MPI), the distribution of magnetic nanoparticles (MNPs) is imaged by moving a field free point (FFP) in space. All MNPs in close vicinity of the FFP contribute to the signal induced on the receive coil. The relaxation behavior of these MNPs are subject to a DC field due to the selection field (SF). In this work, we investigate the effects of the DC field on the relaxation behavior of the MNPs, with the goal of understanding the differences between the measured relaxations in Magnetic Particle Spectrometer (MPS) setups vs. MPI scanners.

## I. Introduction

In Magnetic Particle Imaging (MPI), a field free point (FFP) is created for signal acquisition, which is then moved in space for scanning a targeted region [1]. Only the magnetic nanoparticles (MNPs) in the vicinity of the FFP have unsaturated magnetization and contribute to the signal induced on the receive coil. In practice, the received signal is affected by the relaxation behavior of the MNPs, which causes a loss in signal amplitude as well as a broadening [2]. Previous work has shown that the effective relaxation time constant displays similar trends but at different frequencies in a Magnetic Particle Spectrometer (MPS) setup vs. an MPI scanner [3].

The main difference between these two setups is the absence/presence of the selection field (SF). In an MPI scanner, the signal is not only received from the FFP, but from all MNPs in a relatively small volume in the vicinity of the FFP. These MNPs are subject to a DC field due to the SF, which can alter their relaxation behavior. In return, the effective relaxation of the total measured signal will also be affected. In this study, we investigate the vicinity effects of FFP on the relaxation behavior of MNPs using an in-house MPS setup combined with a DC coil.

## II. Materials and Methods

The relaxation effect is modeled as a convolution of the ideal signal with an exponential relaxation kernel [2].

$$s(t) = s_{ideal}(t) * \left\{\frac{1}{\tau}e^{-\frac{t}{\tau}}u(t)\right\}$$

Here, $\tau$ is the effective relaxation time constant, $u(t)$ is the Heaviside step function, and "$*$" denotes the convolution operation. In this work, the relaxation behavior is investigated using TAURUS (TAU estimation via Recovery of Underlying mirror Symmetry), which does not require any prior information about the MNPs to estimate $\tau$. Accordingly, $\tau$ is computed as follows [4,5].

$$\tau = \frac{S^*_{pos}(f) + S_{neg}(f)}{i2\pi f(S^*_{pos}(f) - S_{neg}(f))}$$

Here, $S_{neg}(f)$ and $S_{pos}(f)$ represent the respective Fourier transforms positive and negative half cycles of $s(t)$. "*" denotes the complex conjugation operation.

An in-house arbitrary waveform MPS setup was used for assessing the relaxation effects. To emulate the vicinity of the FFP of an MPI scanner, a DC Helmholtz coil was designed and implemented. As shown in Fig.1, this DC coil creates a uniform magnetic field along the x-axis, orthogonal to the drive field (DF) of the MPS.

In Figure 2, the simulated and measured sensitivity maps of the DC coil are shown, with 1.76 mT/A measured sensitivity at the center. The measurement chamber of the MPS had 0.7cm diameter and 2cm length, remaining safely within the 95\% homogeneity region of the DC coil. Using a DC power supply (Keysight N8700), this DC coil can generate up to 9mT DC field without any heating issues.

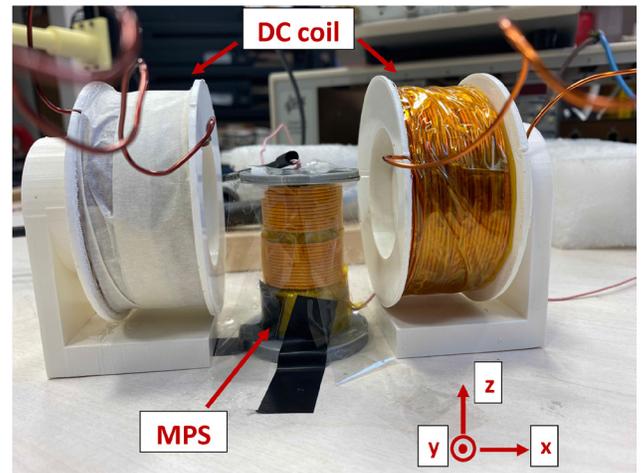

Figure 1: In-house arbitrary waveform MPS setup and the DC coil. The DC coil applies a uniform magnetic field along the x-axis, orthogonal to the drive field of the MPS.

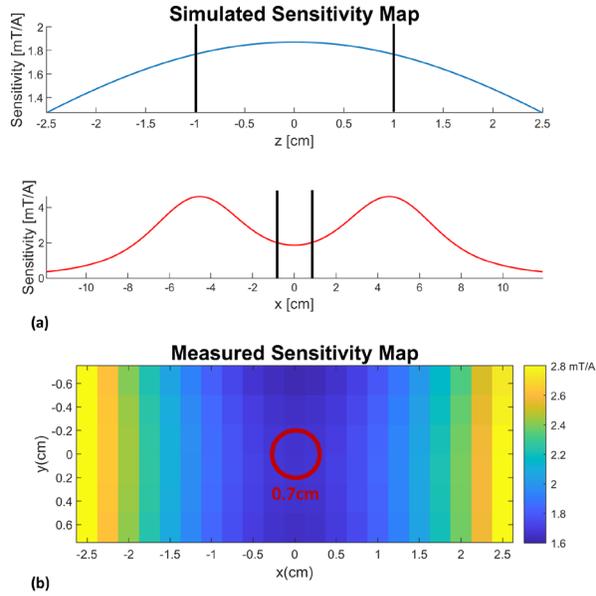

Figure 2: (a) Simulated and (b) measured sensitivity maps of the DC coil. (a) The vertical black lines mark the 95% homogeneity limits, and (b) the red circle marks the central cross-section of the MPS measurement chamber.

For the DF, five different frequencies between 1kHz and 5kHz, and four different amplitudes between 7.5mT and 15mT were applied. A power amplifier (AE Techron 7224) was utilized without the need for impedance matching, thanks to the low inductance of the DF coil. A sample containing 50μL Perimag nanoparticles (Micromod GmbH) diluted with 95μL Deionized (DI) water was utilized. The received signal was amplified using a low-noise preamplifier (SRS SR560).

In total, 660 experiments were performed. At each DF setting, first, measurements with 3 repetitions were performed with the DC coil removed. Then, with the DC coil placed around the MPS setup, 10 different DC fields were applied ranging between 0mT and 9mT, and measurements were performed with 3 repetitions.

## III. Results and Discussion

Figure 3 shows example MNP signals under 3 different DC fields for two different DF settings: at 1kHz and 10mT, and at 5kHz and 10mT. As expected, the signal amplitude decreases with increasing DC field. In addition, the signal becomes wider with increasing DC field, indicating a potential increase in τ for these examples.

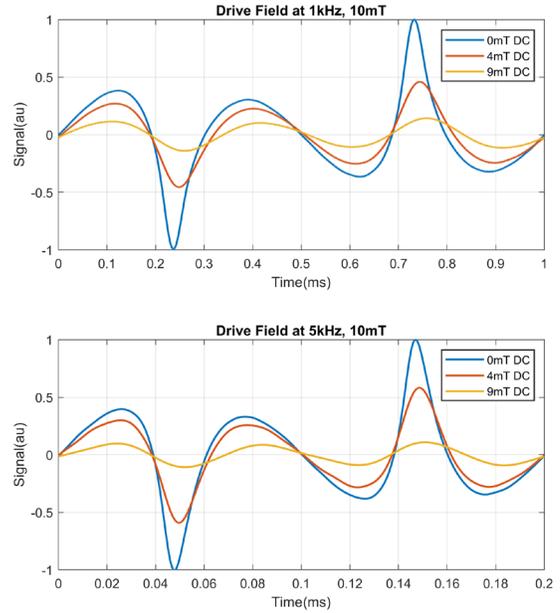

Figure 3: Example MNP signals at two different DF settings and 3 different DC fields.

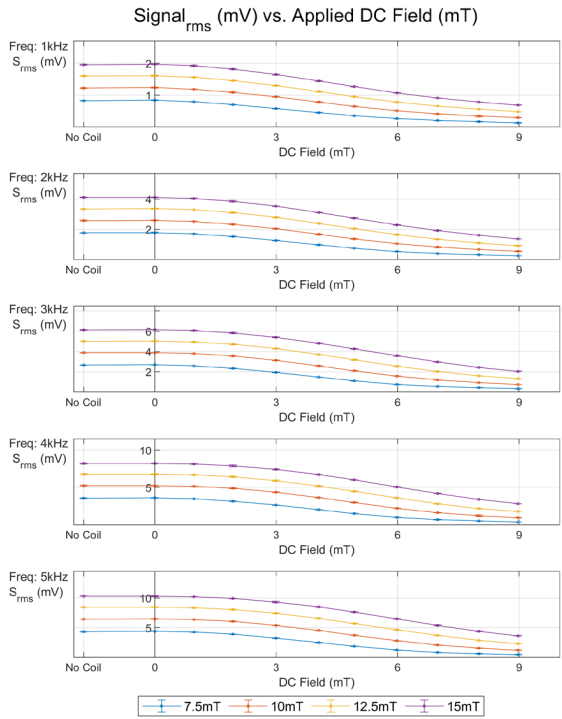

Figure 4: Effects of DC Field on the received RMS signal at 5 different DF frequencies and 4 different DF amplitudes.

Figure 4 shows the root-mean-squared (RMS) signals as a function of the DC field for all DF settings. As the DC field is increased, the signal is reduced due to saturation. Here, the "no coil" case serves as a reference, verifying that the presence of the DC coil without any current does not perturb the MNP signal.

Figure 5 shows τ as a function of the DC field for all DF settings. Again, the "no coil" case is provided as a reference. Overall, τ first decreases and then increases with increasing DC field. The trends in τ at the lowest DF amplitude of 7.5mT slightly diverges from the trends at other DF amplitudes. In all other cases, the applied DC field is smaller than the DF amplitude, whereas at 7.5mT, the DC field is at times comparable to or higher than the DF amplitude. In such a case, MNPs may remain mostly saturated and not rotate sufficiently to align with the DF.

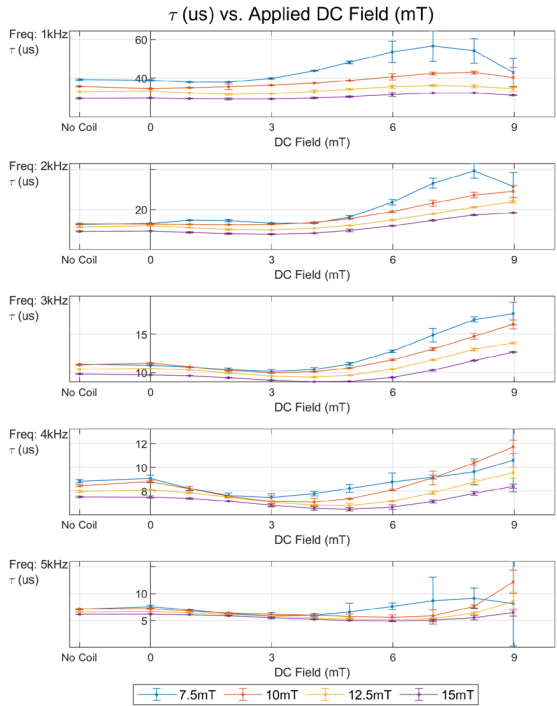

Figure 5: Effects of DC field on τ at 5 different DF frequencies and 4 different DF amplitudes.

A previous study noted that τ gets smaller under DC fields ranging from 0mT to 5mT [3], which is consistent with the results in this work at low DC fields. At large DC fields, however, τ increased monotonically at all frequencies tested. Note that these high DC field cases have relatively small signal levels. Therefore, their contribution to the overall computed τ in the presence of the SF of an MPI scanner may be negligible.

## IV. Conclusion

This work demonstrates the vicinity effects of FFP on the effective relaxation behavior of MNPs. The experiments in our in-house MPS setup combined with a DC coil demonstrate that the effective relaxation time constant first decreases and then increases with increasing DC field. For future work, different MNPs, a wider range of DF settings and DC fields, and different DC field orientations should be tested to better understand the differences in $\tau$ measured in an MPS setup vs. an MPI scanner.

## Author's statement
Conflict of interest: Authors state no conflict of interest.

## Acknowledgments
This work was supported by the Scientific and Technological Research Council of Turkey (Grant No: TUBITAK 120E208).